\documentclass{article}
\usepackage{stmaryrd,amssymb,amsmath,amsthm,url}
\title{On the contribution of backward jumps to instruction sequence expressiveness}

\author{
	Jan A.\ Bergstra    \and
	Inge Bethke \\
\\
  {\small
	  Section Theory of Computer Science,
	  Informatics Institute,
	  University of Amsterdam}\\
	{\small URL: \url{www.science.uva.nl/~{janb,inge}}
	}
}

\newtheorem{theorem}{Theorem}[section]

\newtheorem{proposition}[theorem]{Proposition}

\newtheorem{corollary}[theorem]{Corollary} 
\newtheorem{definition}[theorem]{Definition}   
\newtheorem{example}[theorem]{Example}

\newcommand{\term}{\mathtt{!t}}
\newcommand{\ferm}{\mathtt{!f}}

\newcommand{\Nat}{\ensuremath{\mathbb N}}

\newcommand{\di}{\mathsf{D}}
\newcommand{\st}{\mathsf{S}}

\newcommand{\diamarrow}[1]{\ensuremath{\mathbin{   %
                  \setlength{\unitlength}{1.6ex}
                  \begin{picture}(3,2)
                  \put(.09,.1){$\langle~#1~\rangle$}
                  \put(.65,-.55){\vector(-1,-1){2.5}}
                  \put(2.35,-.55){\vector(1,-1){2.5}}
                  \end{picture}
                  }}}

\newcommand{\prefa}[1]{\ensuremath{\mathbin{        %
                  \setlength{\unitlength}{1.6ex}
                  \begin{picture}(3,2)(-.54,.1)
                  \put(-0.3,.2){$\lbrack~#1~\rbrack$}
                  \put(.95,-.5){\line(0,-1){2.4}}
                  \end{picture}
                  }}}
\newcommand{\prefarrow}[1]{\ensuremath{\mathbin{    %
                  \setlength{\unitlength}{1.6ex}
                  \begin{picture}(3,2)(-.54,.1)
                  \put(-0.3,.2){$\lbrack~#1~\rbrack$}
                  \put(.95,-.5){\vector(0,-1){2.4}}
                  \end{picture}
                  }}}

\newcommand{\tr}{{\mathtt{t}}}
\newcommand{\fa}{{\mathtt{f}}}
\newcommand{\dib}{\mathtt{d}}

\newcommand{\invar}[1]{\mathtt{in}\!\!:\!\! #1}
\newcommand{\aux}[1]{\mathtt{aux}\!\!:\!\! #1}

\newcommand{\derive}[1]{\frac{\partial}{\partial #1}}

\begin{document}

\maketitle

\begin{abstract}
We investigate the expressiveness of backward jumps in a framework of formalized sequential programming called \emph{program algebra}. We show that|if expressiveness is measured in terms of the computability of partial Boolean functions|then backward jumps are superfluous. If we, however, want to prevent explosion of the length of programs, then backward jumps are essential.
\end{abstract}

\section{Introduction}
\label{sec:Intro}
We take the view that sequential programs are in essence instruction sequences which leads to an algebraic approach to the formal description of the semantics of programming languages also known as \emph{program algebra}. It is a framework that permits algebraic reasoning about programs and has been investigated in various settings  (see e.g.\ \cite{BL02,BM84,BM88a,BM88b,W99}). Here the notion of program algebra refers to the concept introduced in \cite{BL02} where the behaviour of a program is taken for a \emph{thread}, i.e.\ a form of process that is tailored  to the description of the behaviour of a deterministic sequential program under execution.

In addition to basic, test and termination instructions, program algebra considers two sorts of unconditional jump instructions: \emph{forward} and \emph{backward} jumps. If only forward jumps are permitted, then threads that perform an infinite sequence of actions are excluded.
In other words, programs for which the execution goes on indefinitely cannot be expressed.
However, in a setting with backward jump instructions also every \emph{regular} infinite thread|i.e.\ every infinite, finite state process|can be described by a finite sequence of primitive instructions.

The aim of this paper is to give an indication of the expressiveness of backward jumps, where expressiveness is measured in terms of the Boolean partial functions that can be computed with the aid of instruction sequences. As it will turn out every partial Boolean function can be computed without backward jumps.  Thus, semantically we can do without backward jumps. However, if we want to avoid an explosion of the length on instruction sequences, then backward jumps are essential.

This paper is organized as follows. Section 2 briefly recalls the program notation PGLB$_{bt}$ and its accompanying thread algebra. In Section 3 we review \emph{services} and the interactions of services with threads. Section 4 investigates the expressiveness of backward jumps.

\section{Instruction sequences and regular threads}
\label{sec:Pre}
In this section, we briefly recall the program notation PGLB$_\text{bt}$ and its accompanying thread algebra. PGLB is a notation for instruction sequences and belongs to a hierarchy of program notations in the program algebra PGA introduced in \cite{BL02} (see also \cite{PZ06}). PGLB$_\text{bt}$ is PGLB with the termination instruction ! refined into two Boolean termination instructions $\term, \ferm$ (see also \cite{BM09a,BM09b,BM09c}). Both PGLB and PGLB$_\text{bt}$ are close to existing assembly languages and have relative jump instructions.

Assume $A$ is a set of constants with typical elements 
$\mathtt{a,b,c,\ldots}$.
PGLB$_\text{bt}(A)$ instruction sequences are then of the following form ($\mathtt{a}\in A$, $l\in\Nat$):
\[I ::= \mathtt{a}\mid +\mathtt{a}\mid -\mathtt{a}
\mid \#l\mid \backslash \# l\mid\term\mid\ferm \mid I;I.\]
The first seven forms above are called  
\emph{primitive instructions}. These are

\begin{description}
\item[1.\hspace{0.4cm}]
\emph{basic instructions} $\mathtt a$ which
prescribe actions that are considered
indivisible and executable in finite time, and which return upon execution a Boolean reply value,
\item[2.-3.]
\emph{test instructions} obtained from basic instructions by prefixing them with either  $+$ 
(positive test instruction) or  $-$ (negative test instruction) which
control subsequent 
execution via the reply of their execution,
\item[4.-5.]
\emph{jump instructions} $\#l, \backslash \#l$ which
prescribe to jump $l$ 
instructions forward and backward, respectively|if possible; otherwise deadlock
occurs|and generate no observable
behavior, and
\item[6.-7.]
 the
\emph{termination instructions} $\term , \ferm$ which prescribe successful
termination and in doing so deliver the Boolean value $\tr$ and $\fa$, respectively.
\end{description}

Complex instruction sequences  are obtained from primitive  instructions using \emph{concatenation}: if
$I$ and $J$ are instruction sequences, then so is 
\[I;J\] 
which is the instruction sequence that lists 
$J$'s primitive instructions right after those of $I$. 
We denote by $\mathcal{IS}(A)$ the set of PGLB$_\text{bt}(A)$ instruction sequences. 

Thread algebra is the behavioural semantics for PGA and was
introduced in e.g.\  \cite{BB03,BL02}
under the name Polarized Process Algebra.

In the setting of PGLB$_\text{bt}(A)$, finite threads are defined inductively by:
\begin{eqnarray*}
\st +&-&\text{the termination thread with positive reply,}\\
\st -&-&\text{the termination thread with negative reply,}\\
\di\hspace{0.09 in}&-&\text{\emph{inaction} or \emph{deadlock}, the inactive thread},\\
T\unlhd\mathtt a \unrhd T'&-&
\text{the \emph{postconditional
composition} of $T$ and $T'$ for action $\mathtt a $,}\\
&&\text{where $T$ and $T'$ are finite threads and $\mathtt a\in A$.}
\end{eqnarray*}
The behaviour of the thread
$T\unlhd\mathtt a \unrhd T'$ starts with the \emph{action} $\mathtt
a$ and continues as $T$ upon reply $\tr$ to $\mathtt a$, and as $T'$
upon reply $\fa$. Note that finite threads always end in $\st +, \st -$ or
$\di$.
We use \emph{action prefix} $\mathtt a \circ T$ as an abbreviation for
$T\unlhd\mathtt a \unrhd T$ and take $\circ$ to bind strongest.

Infinite threads are obtained by guarded recursion. A \emph{guarded recursive specification}  is a set of recursion equations $E=\{E_i=T_i \mid i\in I \}$ where each $T_i$ is of the form $\st +, \st -, \di$ or $T\unlhd\mathtt a \unrhd T'$ with $T,T'$ process terms with variables from $\{E_i\mid i\in I\}$. A \emph{regular} thread is a finite state thread in which infinite paths may occur. Regular threads correspond to finite guarded recursive specifications, i.e.\ guarded recursive specifications with a finite number of recursive equations. To reason about infinite threads, we assume the \emph{Approximation Induction Principle} 
\[
\bigwedge _{n\geq 0} \pi_n(T)=\pi_n(T') \Rightarrow T=T'\ \ \ \ \ (\mathit{AIP}).
\]
\emph{AIP} identifies two threads if their approximations up to any finite depth are identical. The approximation up to depth $n$ of a thread is obtained by cutting it off  after $n$ performed actions. In \emph{AIP}, the approximation up to depth $n$ is phrased in terms of the  \emph{projection} operator $\pi_n$ which is defined  by
\begin{enumerate}
\item $\pi_0(T)=\di$,
\item $\pi_{n+1}(\st +)=\st +$,
\item $\pi_{n+1}(\st -)=\st -$,
\item $\pi_{n+1}(\di)=\di$, and
\item $\pi_{n+1}(T\unlhd\mathtt a \unrhd T')=\pi_n(T)\unlhd\mathtt a \unrhd \pi_n(T')$
\end{enumerate}
for  $n\in \Nat$. 
Every infinite thread $T$ can be identified with its \emph{projective sequence} $(\pi_n(T))_{n\in \Nat}$.

Upon its execution, a basic or test
instruction yields the equally named action in a post
conditional composition.
Thread extraction on PGLB$_\text{bt}(A)$, notation 
\(|X|\)
with $X\in \mathcal{IS}(A)$,
is defined by 
\[ 
|X|=|1,X|
\]
where $|\ ,\ |$ in turn is defined by the equations given in Table~\ref{threadextraction}.
In particular, note that upon the execution of a positive
test instruction $+\mathtt a$, the reply $\tr$ to $\mathtt a$
prescribes to continue with
the next instruction and $\fa$ to skip the next instruction
and to continue with the instruction thereafter; if no such 
instruction is available, deadlock occurs.  For the execution
of a negative
test instruction $-\mathtt a$, subsequent execution
is prescribed by the complementary replies.
\begin{table}[htbp]
\hrule
\[
\begin{array}{rcll}
\\
|i,u_1; \ldots ; u_k|&=&\di & \text {if $i=0$ or $k<i$}\\
|i,u_1; \ldots ; u_k|&=&\mathtt{a}\circ |i+1,u_1; \ldots ; u_k|&\text{if $u_i=\mathtt a$}\\
|i,u_1; \ldots ; u_k|&= &|i+1,u_1; \ldots ; u_k|\unlhd \mathtt a \unrhd |i+2,u_1; \ldots ; u_k|& \text{if $u_i=+ \mathtt a$}\\
|i,u_1; \ldots ; u_k|&=& |i+2,u_1; \ldots ; u_k|\unlhd \mathtt a \unrhd |i+1,u_1; \ldots ; u_k|& \text{if $u_i=- \mathtt a$}\\
|i,u_1; \ldots ; u_k|&=&|i+l,u_1; \ldots ; u_k|& \text{if $u_i=\#l$}\\
|i,u_1; \ldots ; u_k|&=& |i - l,u_1; \ldots ; u_k| & \text{if $u_i=\backslash \#l$ and $i> l$}\\
|i,u_1; \ldots ; u_k|&=& |0,u_1; \ldots ; u_k| & \text{if $u_i=\backslash \#l$ and $i \leq l$}\\
|i,u_1; \ldots ; u_k|&=& \st + &\text{if $u_i = \term$}\\
|i,u_1; \ldots ; u_k|&=& \st - &\text{if $u_i = \ferm$}\\
\\
\end{array}
\]
\hrule
\caption{Equations for thread extraction,
where $\mathtt a$ ranges over the basic instructions and $i,k,l\in\Nat$}
\label{threadextraction}
\end{table}

If we add the rule 
\[
|i, ,u_1; \ldots ; u_k| = \di \text{ if $u_i$ is the beginning of an infinite jump chain}
\]
 then thread extraction on PGLB$_{bt}(A)$ yields regular threads. Conversely, every regular thread corresponds to a PGLB$_{bt}(A)$ instruction sequence after thread extraction.
\begin{example}
{\rm 
We consider the PGLB$_{bt}(A)$ instruction sequence \[X=\mathtt{a;+ b; \#2; \#3; c ;\backslash \# 4; +d; \term; \ferm}.\] Thread extraction of $X$ yields the regular thread
\begin{align*}
E_0&=\mathtt a\circ E_1\\
E_1&=\mathtt c\circ E_1\unlhd \mathtt b\unrhd({\st +}\unlhd \mathtt d\unrhd{ \st -})
\end{align*}
A picture of this thread is

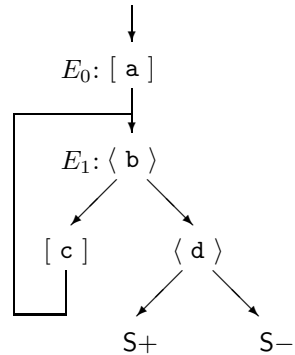
\setlength{\unitlength}{1.6ex}
\begin{picture}(30,20)(-20,0)
\put(1.6,10){$E_1$:}
\put(1.6,15){$E_0$:}
\put(4,10){\diamarrow{\mathtt{b}}}
\put(4,15){\prefarrow{\mathtt{a}}}
\put(7.6,5){\diamarrow{\mathtt{d}}}
\put(0.4,5){\prefa{\mathtt{c}}}
\put(5,0){$\st +$}
\put(5.5,19){\vector(0,-1){2}}
\put(12.5,0){$\st -$}
\put(-1,2){\line(1,0){2.9}}
\put(-1,2){\line(0,1){11}}
\put(-1,13){\line(1,0){6.5}}
\end{picture}
\\[4mm]
Here $[\mathtt{a}]$ corresponds to action prefix and  $\langle \mathtt{a}\rangle$ to postconditional composition with a left hand vector continuing the path in case of a positive reply and a right hand vector in case of a negative reply.

This thread can also given by the projective sequence $(\pi_n(E_0))_{n\in \Nat}$ where
\[
\begin{array}{rcl}
\pi_0(E_0)&=& \di\\
\pi_1(E_0)&=&  \mathtt a\circ\di\\
\pi_2(E_0)&=&\mathtt a\circ \mathtt b \circ \di\\
\pi_3(E_0)&=&\mathtt a\circ (\mathtt c\circ \di \unlhd \mathtt b \unrhd \mathtt d \circ \di )
\end{array}
\]
and $\pi_{n+4}(E_0)= \mathtt a\circ (\mathtt c\circ \pi_{n+1}(E_1) \unlhd \mathtt b \unrhd ({ \st + }\unlhd \mathtt d \unrhd {\st -}))$ where
\[
\begin{array}{rcl}
\pi_0(E_1)&=& \di\\
\pi_1(E_1)&=&  \mathtt b\circ\di\\
\end{array}
\]
and $\pi_{n+2}(E_1)= \mathtt c\circ \pi_{n}(E_1) \unlhd \mathtt b \unrhd ({ \st + }\unlhd \mathtt d \unrhd {\st -})$.
}
\end{example}

For basic information on thread algebra we refer 
to~\cite{BBP05,PZ06}; more advanced matters, such as an operational
semantics for thread algebra, are discussed
in~\cite{BM07}. 

\section{Services}

Services process certain methods which may involve a change of state, and produce reply values.
In the sequel, we let $\mathcal{M}$ be an arbitrary but fixed set of \emph{methods} and 
$\mathcal{R}=\{\tr,\fa,\dib\}$ be the set of \emph{reply values} with $\dib$ the divergent value which is neither true nor false.

A \emph{service} $\mathbb{S}$ consists of
\begin{enumerate}
\item
a set $S$ of \emph{states} in which the service may be,
\item
an \emph{effect} function $\mathit{eff}: {\mathcal{M} \times S}\rightarrow S$ that gives for each method $m$ and state $s$ the resulting state after processing $m$, 
\item
a \emph{yield} function
$\mathit{yld}:{\mathcal{M} \times S}\rightarrow {\mathcal{R}}$ that gives for each method $m$ and state $s$ the resulting reply after processing $m$, and
\item
an \emph{initial state} $s_0 \in S$
\end{enumerate}
satisfying the condition
\[
(\dag)\ \ \ \ \exists s \in S\  \forall m \in \mathcal{M}\  (\ \mathit{yld}(m,s) = \dib\  \&\ 
   \forall{s' \in S}\ (\ \mathit{yld}(m,s') = \dib \Rightarrow \mathit{eff}(m,s') = s\ )\ ).
\]
Given a service $\mathbb{S} = \langle S,\mathit{eff},\mathit{yld},s_0\rangle$ and a method
$m \in \mathcal{M}$,
\begin{enumerate}
\addtocounter{enumi}{4}
\item
the \emph{derived service of $\mathbb{S}$ after processing $m$}, $\derive{m}\mathbb{S}$, is defined by
\[
\derive{m}\mathbb{S} = \langle S,\mathit{eff},\mathit{yld},\mathit{eff}(m,s_0)\rangle 
\]
\item
the \emph{reply of $\mathbb{S}$ after processing $m$}, $\mathbb{S}(m)$, is defined by
\(
\mathbb{S}(m) = \mathit{yld}(m,s_0).
\)
\end{enumerate}
When a request is made to service $\mathbb{S}$ to process method $m$ then
\begin{enumerate}
\addtocounter{enumi}{6}
\item
if $\mathbb{S}(m) \neq \dib$, then the service processes $m$, produces the
reply $\mathbb{S}(m)$, and proceeds as $\derive{m}\mathbb{S}$, but
\item
if $\mathbb{S}(m) = \dib$, then the service rejects the request and proceeds as
a service that rejects any request to process a method.
\end{enumerate}

An \emph{empty} service $\mathbb{S}$ is a service that is unable to process any method, i.e.\
$\mathbb{S}(m)=\dib$ for all $m \in \mathcal{M}$. Given $(\dag)$, we can identify all empty services
and denote it $\delta$. A set  of services is called \emph{closed} if it contains the empty service and is closed under $\derive{m}$ for all $m\in \mathcal{M}$.
\begin{example}\label{boolean_registers}
{\rm
Given the set of methods $\mathcal{M}=\{\mathtt{{set\!\!:\!\!t}}, \mathtt{{set\!\!:\!\!f}}, \mathtt{get}\}$, we consider the set of services $\mathcal{B}=\{B(x)\mid x \in \mathcal{R}\}$ of Boolean registers with initial values $\tr$, $\fa$ and $\dib$, respectively. Here for $x\in \mathcal{R}$, $B(x)=\langle \mathcal{R}, \mathit{eff}, \mathit{yld}, x\rangle$ where
\[
\mathit{eff}(\mathtt{{set\!\!:\!\!t}}, x)=
\begin{cases}
\tr & \text{ if $x =\fa$, and}\\
x &\text{ otherwise}
\end{cases}
\]
\[
\mathit{eff}(\mathtt{{set\!\!:\!\!f}}, x)=
\begin{cases}
\fa & \text{ if $x =\tr$, and}\\
x &\text{ otherwise},
\end{cases}
\]
and $\mathit{eff}(\mathtt{get}, x)=x$; for $m\in \mathcal{M}$, $\mathit{yld}(m,x)=\tr$ if $x\in \{\tr, \fa\}$ and 
$\mathit{yld}(m,\dib)=\dib$. Observe that $\mathcal{B}$ is closed with $\delta = B(\dib)$ and
\[
\begin{array}{ccc}
\derive{\mathtt{set:t}}B(\tr)=B(\tr), &\derive{\mathtt{set:t}}B(\fa)=B(\tr), &\derive{\mathtt{set:t}}B(\dib)=B(\dib), \\[4mm]
\derive{\mathtt{{set:f}}}B(\tr)=B(\fa), &\derive{\mathtt{{set:f}}}B(\fa)=B(\fa), &\derive{\mathtt{{set:f}}}B(\dib)=B(\dib), 
\end{array}
\]
and $\derive{\mathtt{get}}B(x)=B(x)$ for $x\in \mathcal{R}$.
}
\end{example}
A \emph{service family} is a set of services uniquely named by a fixed but arbitray set $\mathcal{F}$ of \emph{foci}.
$\emptyset$ denotes the empty service family, and for $f\in \mathcal{F}$ and service $\mathbb{S}$, $f.\mathbb{S}$ denotes the singleton service family consisting of the named service $f.\mathbb{S}$. $\oplus$ denotes the 
binary composition operator which forms the union of service families under the provision that  named services with the same name collapse to the empty service with that name. For $F\subseteq \mathcal{F}$, $\partial_F$ denotes the unary encapsulation operator which removes the named services with a name in $F$ from a given service family. The axioms for service families  are given in Table \ref{sfaxioms}. 
\begin{table}[htbp]
\[
\begin{array}{lllll}
\hline\\
u \oplus \emptyset = u                                 & \mathrm{SFC1}&\partial_{F}(\emptyset) = \emptyset                       & & \mathrm{SFE1} \\[1mm]
u \oplus v = v \oplus u                              & \mathrm{SFC2}&\partial_{F}(f.\mathbb{S}) = \emptyset & \text{ if $f \in F$}               & \mathrm{SFE2}  \\[1mm]
(u \oplus v) \oplus w = u \oplus (v \oplus w)      & \mathrm{SFC3} &\partial_{F}(f.\mathbb{S}) = f.\mathbb{S}      & \text{ if $f\not\in F$}            & \mathrm{SFE3} \\[1mm]
f.\mathbb{S }\oplus f.\mathbb{S}' = f.\delta                       & \mathrm{SFC4}&
\partial_{F}(u \oplus v) =
\partial_{F}(u) \oplus \partial_{F}(v)                    & & \mathrm{SFE4}\\
\\
\hline
\end{array}
\]
\caption{Axioms for binary composition and encapsulation of service families with $f\in \mathcal{F}$, $F\subseteq \mathcal{F}$ and services $\mathbb{S}, \mathbb{S}'$. \label{sfaxioms}}
\end{table}

Let $A=\{f.m \mid f\in \mathcal{F},\ m\in \mathcal{M}\}\cup \{\mathtt{tau}\}$ where $\mathtt{tau}$ denotes a basic internal action which does not have any side effects and always replies true. Then a thread may make \emph{use} of  services by performing a basic action for the purpose of
requesting a named service to process a method and to return a \emph{reply}
value at completion of the processing of the method.
In the sequel, we combine threads with services  and extend the
combination with the two operators $/$ and $!$ which relate to this kind of interaction
between threads and services.

The thread denoted by a closed term of the form $T / \mathcal{S}$ is
the thread that results from processing
the method of each basic action with a focus of the service family
denoted by $\mathcal{S}$ that the thread denoted by $T$ performs, where the
processing is done by the service in that service family with the focus
of the basic action as its name.
When the method of a basic action performed by a thread is processed by
a service, the service changes in accordance with the method concerned,
and affects the thread as follows: the basic action turns into the
internal action $\mathtt{tau}$ and the two ways to proceed reduce to one on the
basis of the reply value produced by the service.
The value denoted by a closed term of the form $T ! \mathcal{S}$ is the
Boolean value that the thread denoted by $T / \mathcal{S}$ delivers at its
termination,  and the
value $\dib$ if it does not terminate.
The axioms for the use and the reply operator were first given in \cite{BM09b} and are
listed in Tables~\ref{axioms-use}, 
and~\ref{axioms-reply}. In their original version, the axiomatizations contain the axioms U3 and R3 concerning the use and reply of the unpolarized termination $\st$ with service families. Since we only consider boolean termination, we have omitted these axioms.

\begin{table}[hbtp]
\[
\begin{array}{lll}
\hline \\
{\st+}/ u = \st+                             & & \mathrm{U1} \\[1mm]
{\st-}/ u = \st-                              & & \mathrm{U2} \\[1mm]
\di / u = \di                           & & \mathrm{U4} \\[1mm]
(\mathtt{tau} \circ x) / u = \mathtt{tau} \circ (x / u)      & & \mathrm{U5} \\[1mm]
(x\unlhd f.m\unrhd y) / \partial_{\{f\}}(u) =
(x / \partial_{\{f\}}(u))\unlhd f.m\unrhd (y / \partial_{\{f\}}(u))
                                                       & & \mathrm{U6} \\[1mm]
(x\unlhd f.m\unrhd y) / (f.\mathbb{S} \oplus \partial_{\{f\}}(u)) =
\mathtt{tau} \circ (x / (f.\derive{m}\mathbb{S} \oplus \partial_{\{f\}}(u)))
                                    & \text{ if $\mathbb{S}(m) = \tr$} & \mathrm{U7} \\[1mm]
(x\unlhd f.m \unrhd y) / (f.\mathbb{S} \oplus \partial_{\{f\}}(u)) =
\mathtt{tau} \circ (y / (f.\derive{m}\mathbb{S} \oplus \partial_{\{f\}}(u)))
                                    & \text{ if $\mathbb{S}(m) = \fa$} & \mathrm{U8} \\[1mm]
(x\unlhd f.m \unrhd y) / (f.\mathbb{S} \oplus \partial_{\{f\}}(u)) = \di
                                    & \text{ if $\mathbb{S}(m) = \dib$}   & \mathrm{U9} \\[1mm]
\\
\hline
\end{array}
\]
\caption{Axioms for the use operator with $f\in \mathcal{F}$, $m\in \mathcal{M}$ and service $\mathbb{S}$ \label{axioms-use}}
\end{table}

\begin{table}[hbtp]
\[
\begin{array}{lll}
\hline \\
{\st+}!u = \tr                           & & \mathrm{R1} \\[1mm]
{\st-}!u = \fa                            & & \mathrm{R2} \\[1mm]
\di !u = \dib                           & & \mathrm{R4} \\[1mm]
(\mathtt{tau} \circ x) !u = x ! u      & & \mathrm{R5} \\[1mm]
(x\unlhd f.m\unrhd y) ! \partial_{\{f\}}(u) =\dib
                                                       & & \mathrm{R6} \\[1mm]
(x\unlhd f.m\unrhd y) ! (f.\mathbb{S} \oplus \partial_{\{f\}}(u)) =
x !(f.\derive{m}\mathbb{S} \oplus \partial_{\{f\}}(u))
                                    & \text{ if $\mathbb{S}(m) = \tr$} & \mathrm{R7} \\[1mm]
(x\unlhd f.m \unrhd y) !(f.\mathbb{S} \oplus \partial_{\{f\}}(u)) =
y ! (f.\derive{m}\mathbb{S} \oplus \partial_{\{f\}}(u))
                                    & \text{ if $\mathbb{S}(m) = \fa$} & \mathrm{R8} \\[1mm]
(x\unlhd f.m \unrhd y) ! (f.\mathbb{S} \oplus \partial_{\{f\}}(u)) = \dib
                                    & \text{ if $\mathbb{S}(m) = \dib$}   & \mathrm{R9} \\[1mm]
\\
\hline
\end{array}
\]
\caption{Axioms for the reply operator with $f\in \mathcal{F}$, $m\in \mathcal{M}$ and service $\mathbb{S}$ \label{axioms-reply}}
\end{table}
\begin{example}
{\rm
We continue with Example \ref{boolean_registers} and 
put $\mathcal{F}= \Nat$. We let  $Eq(1,2)$ be the PGLB$_{bt}(A)$ instruction sequence
\[
+1.\mathtt{get};\#2;\#4;+2.\mathtt{get};\term; \ferm; -2.\mathtt{get};\backslash \#3; \backslash \#3
\]
which intuitively describes a finite thread that compares  2 Boolean registers and returns the reply $\tr$ if their values are not divergent and equal, $\fa$ if their values are not divergent but different, and $\dib$ otherwise. Indeed, formalizing this interaction in the setting of services we put $\mathcal{S}=1.B(b_1)\oplus 2.B(b_2)$ and compute
\[
\begin{array}{rcl}
|Eq(1,2)|!\mathcal{S}&=&( ({\st+} \unlhd 2. \mathtt{get} \unrhd {\st-})\unlhd 1.\mathtt{get} \unrhd ({\st-} \unlhd 2. \mathtt{get} \unrhd{ \st+}))!\mathcal{S}\\
\\
&=& \begin{cases}
({\st+} \unlhd 2. \mathtt{get} \unrhd {\st-})!\mathcal{S} & \text{ if $b_1=\tr$,}\\
({\st-} \unlhd 2. \mathtt{get} \unrhd{ \st+})!\mathcal{S} & \text{ if $b_1=\fa$, and}\\
\dib  & \text{ if $b_1=\dib$.}
\end{cases}\\
\\
&=& \begin{cases}
\tr & \text{ if $b_1=b_2\neq \dib$,}\\
\fa& \text{ if $\dib\neq b_1\neq b_2\neq \dib$, and}\\
\dib  & \text{ if $b_1=\dib$ or $b_2=\dib$.}
\end{cases}
\end{array}
\]
We let $E(m,n)$ be the generic equality test for the registers $b_m,b_n$ and 
\[
E(1,2,3)=+1.\mathtt{get};\#2;\#4;-2.\mathtt{get};\ferm; \# 4;+2.\mathtt{get};\backslash \#3;0.\mathtt{set\!\!:\!\!f};  E(0,3)
\]
the  lazy equality test of 3 registers which stores an intermediate result in the auxiliary register $b_0$.
Observe that $|Eq(1,2,3)|/0.B(\tr)$
\[
\begin{array}{rcl}
&=&(( |E(0,3)|\unlhd 2. \mathtt{get} \unrhd {\st-})\unlhd 1.\mathtt{get} \unrhd ({\st-} \unlhd 2. \mathtt{get} \unrhd (0.\mathtt{set\!\!:\!\!f}\circ |E(0,3)|)))/0.B(\tr)\\
\\
&=& \begin{cases}
(|E(0,3)|\unlhd 2. \mathtt{get} \unrhd {\st-})/0.B(\tr)& \text{ if $b_1=\tr$,}\\
({\st-} \unlhd 2. \mathtt{get} \unrhd (0.\mathtt{set\!\!:\!\!f}\circ |E(0,3)|))/0.B(\tr) & \text{ if $b_1=\fa$, }\\
\di  & \text{ if $b_1=\dib$,}
\end{cases}\\
\\
&=& \begin{cases}
|E(0,3)|/0.B(\tr)& \text{ if $b_1=\tr =b_2$,}\\
(0.\mathtt{set\!\!:\!\!f}\circ |E(0,3))|/0.B(\tr)& \text{ if $b_1=\fa =b_2$,}\\
{\st-}& \text{ if $\dib\neq b_1\neq b_2\neq \dib$,}\\
\di  & \text{ if $b_1=\dib$ or $b_2=\dib$,}
\end{cases}
\\
&=& \begin{cases}
|E(0,3)|/0.B(\tr)& \text{ if $b_1=\tr =b_2$,}\\
\mathtt{tau}\circ (|E(0,3)|/0.B(\fa))& \text{ if $b_1=\fa =b_2$,}\\
{\st-}& \text{ if $\dib\neq b_1\neq b_2\neq \dib$,}\\
\di  & \text{ if $b_1=\dib$ or $b_2=\dib$,}
\end{cases}
\\
&=& \begin{cases}
 \mathtt{tau} \circ {\st +}& \text{ if $b_1=\tr =b_2=b_3$,}\\
\mathtt{tau}\circ  {\st -}& \text{ if $b_1=\tr =b_2$ and $b_3=\fa$,}\\
\mathtt{tau}\circ \mathtt{tau} \circ {\st +}& \text{ if $b_1=\fa =b_2=b_3$,}\\
\mathtt{tau}\circ \mathtt{tau} \circ {\st -}& \text{ if $b_1=\fa =b_2$ and $b_3=\tr$,}\\
{\st-}& \text{ if $\dib\neq b_1\neq b_2\neq \dib$, and}\\
\di  & \text{ if $b_1=\dib$ or $b_2=\dib$ or $b_1=b_2\neq \dib =b_3$.}
\end{cases}
\end{array}
\]
Thus
\[
\begin{array}{rcl}
(|E(1,2,3)|/0.B(\tr))! \oplus_{i=1}^{3}B(b_i)&=&
\begin{cases}
\tr& \text{ if $b_1=b_2=b_3\neq \dib$,}\\
\dib  & \text{ if $b_1=\dib$ or $b_2=\dib$ or $b_1=b_2\neq \dib =b_3$,}\\
\fa& \text{ otherwise.}
\end{cases}
\end{array}
\]
Here $\oplus_{i=1}^{3}B(b_i)$ abbreviates the service family $1.B(b_1)\oplus 2.B(b_2)\oplus 3.B(b_3)$.
An equality test for 3 registers can be written in several ways: e.g.\ without backward jumps by replacing the jump $\backslash \#3$ by $\ferm$. Also the use of an auxiliary register can be omitted. We shall come back to this issue in Proposition \ref{without}.
}
\end{example}
In the case of regular threads, one can show that projection distributes over use, i.e. that $\pi_n(T/\mathcal{S})=\pi_n(T)/\mathcal{S}$ for $n\in \Nat$. It then follows from the Approximation Induction Principle that 
\[
\bigwedge_{n \geq 0}\pi_{n}(T)/\mathcal{S} = \pi_{n}(T')/\mathcal{S}' \Rightarrow T /\mathcal{S} = T' / \mathcal{S}'.
\]
For more information about services we refer to \cite{BM09b,BP02}.
\section{Backward jumps}
Backward jumps $\backslash \#l$ ($l\in \Nat$) are of obvious importance for constructing instruction sequences with loops. Now one may ask how vital are backward jumps? Consider $\mathtt{a}; \backslash \#1$|a PGLB$_{bt}(A)$ instruction sequence which prescribes the execution of the atomic action $\mathtt{a}$ followed by a backward jump of length 1. This instruction sequence produces the thread $T$ with $T=\mathtt{a} \circ T$|a thread that performs the action $\mathtt{a}$ followed by a recursive invocation of the thread. Clearly no $X\in \mathcal{IS}(A)$ can produce a thread with an unbounded number of successive $\mathtt{a}$'s. Thus backward jumps add to the expressiveness of PGLB$_{bt}(A)$. 

In the field of expressiveness and computational complexity one classifies computational problems according to their inherent difficulty. A computational problem can be viewed as an infinite collection of instances together with a solution for every instance. It is conventional to represent both instances and solutions by binary strings. We adopt $\mathbb{B}=\{\tr, \fa\}$ as the preferred binary alphabet and associate with each computational problem a partial function $F:\mathbb{B}^* \overset{p}{\longrightarrow}  \mathbb{B}$ deciding partially whether a certain instance has a solution. In this section, we study the complexity of computing such computational problems.
In the sequel, we denote by $\mathcal{IS}^\mathit{lf}(A)$ the set of \emph{loop-free} PGLB$_{bt}(A)$ instruction sequences, i.e.\ the set of PGLB$_{bt}(A)$ instruction sequences without backward jumps. Moreover, we write $\mathit{length}(I)$ for the number of instructions of $I\in \mathcal{IS}(A)$.
\begin{definition}
\mbox{}
\begin{enumerate}
\item Let $\mathcal{F}=\mathcal{F}_{\mathtt{in}} \cup \mathcal{F}_{\mathtt{aux}}$ where $\mathcal{F}_{\mathtt{in}}=
\{\invar{n}\mid n\in \Nat\}$ and $\mathcal{F}_{\mathtt{aux}}=
\{\aux{n}\mid n\in \Nat\}$, $\mathcal{M}=\{\mathtt{{set\!\!:\!\!t}}, \mathtt{{set\!\!:\!\!f}}, \mathtt{get}\}$ and $A=\{f.m \mid f\in \mathcal{F}, m\in \mathcal{M}\}$.
\item Let $f_1, \ldots , f_n\in \mathcal{F}$ and $\mathbb{S}_1, \ldots , \mathbb{S}_n$ be services. Then $\oplus_{i=1}^n f_i.\mathbb{S}_i$ denotes the service family $f_1.\mathbb{S}_1 \oplus \cdots \oplus f_n.\mathbb{S}_n$.
\item Let $F:\mathbb{B}^k \overset{p}{\longrightarrow} \mathbb{B}$ be a $k$-ary partial function on the Booleans $\mathbb{B}$.   $I\in \mathcal{IS}(A)$ is said to \emph{compute} $F$ \emph{using $l$ auxiliary registers} if for all $b_1, \ldots , b_k \in \mathbb{B}$
\[
(|I|/\oplus_{i=1}^{l} \aux{i}.B(\tr))! \oplus_{i=1}^{k}\invar{i}.B(b_i)= 
\begin{cases}
F(b_1, \ldots , b_k) &\text{ if $F(b_1, \ldots , b_k)$ is defined},\\
\dib & \text{ otherwise.}
\end{cases}
\]
Moreover, we say that \emph{$I$ computes $F$} if $I$ computes $F$ using $l$ auxiliary registers for some $l\in \Nat$, and \emph{$I$ computes $F$ without the use of auxiliary registers} if $l=0$. 
\end{enumerate}
\end{definition}
\begin{proposition}\label{without}
Let $F:\mathbb{B}^k \overset{p}{\longrightarrow} \mathbb{B}$ be a $k$-ary partial function on the Booleans $\mathbb{B}$. Then $F$ can be computed by an $I\in \mathcal{IS}^\mathit{lf}(A)$ with length $3\times 2^k-2$ without the use of auxiliary registers. 
\end{proposition}
{\bf Proof:} By induction on $k$, we construct an instruction sequence $I_F\in \mathcal{IS}^\mathit{lf}(A)$ that computes $F$. If $k=0$, then $F()$ is either $\tr$, $\fa$ or undefined.  Thus we can take for $I_F$ either $\term$, $\ferm$ or $\#0$. Let $F$ be $k+1$-ary and consider the functions $G_b(b_1, \ldots ,b_k)=F(b_1, \ldots ,b_k, b)$ with $b\in \{\tr, \fa\}$. By the induction hypothesis $G_b$ can be  computed by some
$I_{G_b}\in \mathcal{IS}^\mathit{lf}(A)$ with length $3\times 2^k-2$ without the use of auxiliary registers.  Then
\[
(|-\invar{k+1}.\mathtt{get};\#3\times 2^k-1; I_{G_\tr};I_{G_\fa}|/\emptyset)!\oplus_{i=1}^{k+1} i.B(b_i)=
\begin{cases}
(|I_{G_\tr}|/\emptyset)!\oplus_{i=1}^{k} i.B(b_i) & \text{ if $b_{k+1}=\tr$,}\\
(|I_{G_\fa}|/\emptyset)!\oplus_{i=1}^{k} i.B(b_i) & \text{ if $b_{k+1}=\fa$, }\\
\dib & \text{ otherwise}.
\end{cases}
\]
Thus $I_F= -\invar{k+1}.\mathtt{get};\#3\times 2^k-1; I_{G_\tr};I_{G_\fa}$ computes $F$ without the use of auxiliary registers or backward jumps and has length $2+2\times (3\times 2^k -2)=3\times 2^{k+1}-2$. \hfill $\Box$

Thus backward jumps are not necessary for the computation of partial Boolean functions. However, they can make a contribution to the  expressiveness of PGLB$_{bt}(A)$ by allowing shorter instruction sequences for computing a given computational problem.
\begin{definition}
For $F:\mathbb{B}^* \overset{p}{\longrightarrow} \mathbb{B}$, we denote by $F_k$ ($k\in \Nat$) the restriction of $F$ to $\mathbb{B}^k$ and
distinguish the following three classes of computational problems.
\begin{enumerate}
\item $\mathcal{IS}^\mathit{lf}_P(A)=$
\[
\begin{array}{rcl}
\{F: \mathbb{B}^* \overset{p}{\longrightarrow} \mathbb{B} & \mid & \text{ there exists a polynomial function $h:\Nat \rightarrow \Nat$}\\
&& \text{ such that  for all $k\in \Nat$,}\\
&& \text{ $F_k$ can be computed by an $I\in \mathcal{IS}^\mathit{lf}(A)$ with $\mathit{length}(I)\leq h(k)$}\}
\end{array}
\]
\item $\mathcal{IS}_P(A)=$
\[
\begin{array}{rcl}
\{F: \mathbb{B}^* \overset{p}{\longrightarrow} \mathbb{B} & \mid & \text{ there exists a polynomial function $h:\Nat \rightarrow \Nat$}\\
&& \text{ such that  for all $k\in \Nat$,}\\
&& \text{ $F_k$ can be computed by an $I\in \mathcal{IS}(A)$ with $\mathit{length}(I)\leq h(k)$}\}
\end{array}
\]
\item $\mathcal{IS}^\mathit{lf}_E(A)=$
\[
\begin{array}{rcl}
\{F: \mathbb{B}^* \overset{p}{\longrightarrow} \mathbb{B} & \mid & \text{ there exists a $c\in \Nat$ such that  for all $k\in \Nat$,}\\
&& \text{ $F_k$ can be computed by an $I\in \mathcal{IS}^\mathit{lf}(A)$ with $\mathit{length}(I)\leq c\times 2^k$}\}
\end{array}
\]
\end{enumerate}
\end{definition}
In the sequel we denote by $[\mathbb{B}^* \overset{p}{\longrightarrow} \mathbb{B} ]$ the set of all partial functions from $\mathbb{B}^*$ to $\mathbb{B}$, and by
$[\mathbb{B}^* \longrightarrow \mathbb{B} ]$ the set of all total functions from $\mathbb{B}^*$ to $\mathbb{B}$.
Restating Proposition \ref{without}, we have 
\begin{proposition}\label{boven}
\(
\mathcal{IS}^\mathit{lf}_E(A)=[\mathbb{B}^* \overset{p}{\longrightarrow} \mathbb{B} ]
\)
\end{proposition}
In nonuniform complexity theory, P/poly is the complexity class of computational problems solved by a polynomial-time Turing machine with a polynomial-bounded advice function. It is also equivalently defined as the class PSIZE of problems that have polynomial-size Boolean circuits.
\begin{theorem}\label{beneden}
\(\mathcal{IS}^\mathit{lf}_P(A)\cap [\mathbb{B}^* \longrightarrow \mathbb{B}] = \mathrm{P/poly}\)
\end{theorem}
{\bf Proof:} 
We shall prove the inclusion $\subseteq$ using the definition of P/poly in terms of Turing machines that take advice, and the inclusion $\supseteq$ using the definition in terms of Boolean circuits.

$\subseteq$: Suppose that $F\in \mathcal{IS}^\mathit{lf}_P(A)\cap [\mathbb{B}^*\rightarrow \mathbb{B}]$. Then, for all $k\in \Nat$, there exists an $I_k \in \mathcal{IS}^\mathit{lf}(A)$ that computes $F_k$ with $\mathit{length}(I_k)$ polynomial in $k$. Then $F$ can be computed  by a Turing machine that, on input of size $k$, takes a binary description of $I_k$ as advice and then just simulates the execution of $I_k$. It is easy to see that under the assumption that instructions of the form $\invar{i}.m, +\invar{i}.m, -\invar{i}.m$ with $i>k$,  and $\aux{i}.m, +\aux{i}.m, -\aux{i}.m$, and $\#i$ with $i>\mathit{length}(I_k)$ do not occur in $I_k$, the size of the description of $I_k$ and the number of steps that it takes to simulate its execution are both polynomial in $k$. It is obvious that we can make the assumption without loss of generality. Hence, $F$ is also in P/poly. 

$\supseteq$: We first show that a  function $F: \mathbb{B}^k \rightarrow \mathbb{B}$ that is induced by a Boolean circuit $C$ consisting of   
  NOT, AND and OR gates  can be computed by an $I_C\in \mathcal{IS}^{lf}(A)$. More precisely,  assuming that $\{g_{i_1}, \ldots , g_{i_n}\}$ 
  ($i_1, \ldots , i_n \in \Nat$) is a topological ordering of the gates with output node $g_{i_n}$, we prove by induction on  $n$ that we may assume that $I_C$ is of the form $I;+\aux{i_n}.\mathtt{get}; \term;\ferm$ for some $I\in \mathcal{IS}^{lf}(A)$ with $\mathit{length}(I)\leq 4\times n$.

If $n=1$, then depending on the form of the single gate either
\[
\begin{array}{rcll}
I_\neg &= &+\invar{i}.\mathtt{get};\aux{i_1}.\mathtt{set\!\!:\!\!f};+\aux{i_1}.\mathtt{get}; \term;\ferm,&\\
I_\wedge&=& -\invar{i}.\mathtt{get}; \#2; -\invar{j}.\mathtt{get}; \aux{i_1}.\mathtt{set\!\!:\!\!f};+\aux{i_1}.\mathtt{get}; \term;\ferm, \text{ or}\\
I_\vee&=& +\invar{i}.\mathtt{get}; \#3; -\invar{j}.\mathtt{get}; \aux{i_1}.\mathtt{set\!\!:\!\!f};+\aux{i_1}.\mathtt{get}; \term;\ferm
\end{array}
\]
 with properly chosen $i,j$ comply.  For the induction step we again have to distinguish three cases. We here consider only the case that $g_{i_n}$ is an $AND$ gate. Suppose that the input of $g_{i_n}$ are the  output gates $g_{i_l}$ and $g_{i_m}$ of the subcircuits $C'$ and $C''$. By the induction hypothesis we may assume that the functions induced by $C'$ and $C''$ can be computed by the $\mathcal{IS}^{lf}(A)$ instruction sequences $I_{C'}=I'; +\aux{i_l}.\mathtt{get}; \term;\ferm$ and $I_{C''}=I''; +\aux{i_m}.\mathtt{get}; \term;\ferm$ with
$\mathit{length}(I_{C'})\leq 4\times |C'|$ and $\mathit{length}(I_{C''})\leq 4\times |C''|$ where the sizes $|C'|$ and $|C''|$ are the number of gates in the respective subcircuits. Then
\[
I_C=I';I'';-\aux{i_l}.\mathtt{get}; \#2; -\aux{i_m}.\mathtt{get}; \aux{i_n}.\mathtt{set\!\!:\!\!f};+\aux{i_n}.\mathtt{get}; \term;\ferm
\]
computes $F$ and $\mathit{length}(I)=\mathit{length}(I')  + \mathit{length}(I'')\leq 4\times |C'| + 4\times |C'|\leq 4\times n$. If one input is an input node, a shorter instruction sequence suffices, e.g.\ $I';-\invar{j}.\mathtt{get}; \#2; -\aux{i_l}.\mathtt{get}; \aux{i_n}.\mathtt{set\!\!:\!\!f};+\aux{i_n}.\mathtt{get}; \term;\ferm$.

Now suppose that $F\in \text{P/poly}$. Then, for all $k\in \Nat$, there exists a Boolean circuit $C_k$ such that $C_k$ computes $F_k$ and the size of $C_k$ is polynomial in $k$. From the above and the fact that linear in the size of $C_k$ implies polynomial in $k$, it follows that $F$ is also in $\mathcal{IS}^\mathit{lf}_P(A)$. \hfill $\Box$

Combining Proposition \ref{boven} and the previous theorem, we have
\begin{corollary}\label{hierarchy}
\(\mathrm{P/poly} \subsetneq \mathcal{IS}^\mathit{lf}_P(A) \subseteq \mathcal{IS}_P(A) \subseteq \mathcal{IS}^\mathit{lf}_E(A)\)
\end{corollary}
In the remainder of this section we shall show|adopting  a reasonable assumption|that also the second inclusion is proper.

The satisfiability problem $3SAT$ is concerned with efficiently finding a satisfying assignment to a propositional formula. The input  is a conjunctive normal form where each clause is limited to at most 3 literals|a \emph{3-CNF formula}. The goal is to find an assignment to the variables that makes the entire expression true, or to prove that no such assignment exists. This problem is NP-complete, and therefore no polynomial-time algorithm can succeed on all 3-CNF formulae unless $NP\subseteq \mathrm{P/poly}$ \cite{C71,L73}. The latter implies the collapse of the polynomial hierachy as was proved by Karp and Lipton in 1980 \cite{KL80}.

$3SAT(k)$ can be computed by instruction sequences with polynomial length if we allow backward jumps. Under the hypothesis that $NP\not \subseteq \mathrm{P/poly}$, it then follows that instruction sequences for this decision problem without backward jumps have to be significantly longer.
\begin{theorem}
$3SAT \in \mathcal{IS}_P(A)$
\end{theorem}
{\bf Proof:}
If the number of Boolean variables is $k$, then there are $8k^3$ possible clauses of length 3|we allow multiple occurrences of a variable in a single clause and neglect the order of the literals. We will encode a 3-CNF $\psi$ over $k$ Boolean variables as a sequence of Boolean values $\langle b\rangle_\psi$ of length $8k^3$ where a $\tr$ indicates that a certain clause occurs in the 3-CNF and a $\fa$ excludes the clause. Vice versa, given a sequence
$\langle b\rangle\in \mathbb{B}^{8k^3}$ we denote the 3-CNF obtained from $\langle b\rangle$ by $\psi_{\langle b\rangle}$ and define $3SAT(k): \mathbb{B}^{8k^3} \rightarrow \mathbb{B}$ by
\begin{equation*}
3SAT(k)(\langle b\rangle)=
\begin{cases}
\tr & \text{ if $\psi_{\langle b\rangle}$ is satisfiable,}\\
\fa & \text{ otherwise.}
\end{cases}
\end{equation*}
We let $\{v_1, \ldots , v_k\}$ be Boolean variables and define for $\langle l,m,n,i\rangle \in \{1, \ldots, k\}^3\times \{1, \ldots , 8\}$ the clause $\gamma_{\langle l,m,n,i\rangle}$ by
\begin{equation*}
\gamma_{\langle l,m,n,i\rangle}=
\begin{cases}
\begin{array}{cccccl}
v_l& \vee& v_m &\vee& v_n & \text{ if $i=1$,}\\
v_l &\vee& v_m &\vee& \neg v_n & \text{ if $i=2$,}\\
v_l& \vee& \neg v_m &\vee& v_n & \text{ if $i=3$,}\\
v_l &\vee& \neg v_m &\vee&\neg v_n & \text{ if $i=4$,}\\
\neg v_l &\vee& v_m& \vee& v_n & \text{ if $i=5$,}\\
\neg v_l &\vee &v_m& \vee&\neg v_n & \text{ if $i=6$,}\\
\neg v_l&\vee &\neg v_m &\vee& v_n & \text{ if $i=7$,}\\
\neg v_l &\vee& \neg v_m& \vee& \neg v_n & \text{ if $i=8$,}
\end{array}
\end{cases}
\end{equation*}
and the instruction sequence $\mathit{CHECK}_{\langle l,m,n,i\rangle}$ by
\begin{equation*}
\mathit{CHECK}_{\langle l,m,n,i\rangle}=
\begin{cases}
+\aux{l}.\mathtt{get}; \# 2; +\aux{m}.\mathtt{get}; \#2; +\aux{n}.\mathtt{get}&\text{ if $i=1$,}\\
+\aux{l}.\mathtt{get}; \# 2; +\aux{m}.\mathtt{get}; \#2; -\aux{n}.\mathtt{get} & \text{ if $i=2$,}\\
+\aux{l}.\mathtt{get}; \# 2; -\aux{m}.\mathtt{get}; \#2; +\aux{n}.\mathtt{get}& \text{ if $i=3$,}\\
+\aux{l}.\mathtt{get}; \# 2; -\aux{m}.\mathtt{get}; \#2; -\aux{n}.\mathtt{get} & \text{ if $i=4$,}\\
-\aux{l}.\mathtt{get}; \# 2; +\aux{m}.\mathtt{get}; \#2; +\aux{n}.\mathtt{get} & \text{ if $i=5$,}\\
-\aux{l}.\mathtt{get}; \# 2; +\aux{m}.\mathtt{get}; \#2; -\aux{n}.\mathtt{get} & \text{ if $i=6$,}\\
-\aux{l}.\mathtt{get}; \# 2; -\aux{m}.\mathtt{get}; \#2; +\aux{n}.\mathtt{get} & \text{ if $i=7$,}\\
-\aux{l}.\mathtt{get}; \# 2; -\aux{m}.\mathtt{get}; \#2; -\aux{n}.\mathtt{get}& \text{ if $i=8$.}
\end{cases}
\end{equation*}
Observe that the snippet $\mathit{CHECK}_{\langle l,m,n,i\rangle}$ checks whether a certain assignment|held in the auxiliary registers 
$b_1 , \ldots , b_k$|satisfies clause $\gamma_{\langle l,m,n,i\rangle}$.  We fix an arbitrary bijection $\phi:\{1, \ldots, 8n^3\} \rightarrow \{1, \ldots, n\}^3\times \{1, \ldots , 8\}$ and put 
\[
m\rightarrow \mathit{CHECK}_{\phi(m)}= -\invar{m}.\mathtt{get};\#8;\mathit{CHECK}_{\phi(m)};\#2;\#9
\]
if $1\leq m < 8k^3$, and 
\[
8k^3\rightarrow \mathit{CHECK}_{\phi(8k^3)}= -\invar{8k^3}.\mathtt{get};\#6;\mathit{CHECK}_{\phi(8k^3)};\term
\]
and join the conditional checks to form the instruction sequence
\[
\mathit{CHECK}= 1\rightarrow \mathit{CHECK}_{\phi(1)}; \ldots ; 8k^3\rightarrow \mathit{CHECK}_{\phi(8k^3)}.
\]
Thus, if $\gamma_{\phi(m)}$ is a clause of the 3-CNF that is satisfied by the current assignment, or if the clause does not occur in the 3-CNF, then 
execution of $\mathit{CHECK}$ continues after the snippet $m\rightarrow \mathit{CHECK}_m$ with the snippet $m+1 \rightarrow \mathit{CHECK}_{m+1}$ if $m< 8k^3$ and  terminates with reply $\tr$ if $m=8k^3$. If, however,  $\gamma_{\phi(m)}$ is not satisfied by the assignment then execution jumps with chained jumps of length 9 or|if $m=8k^3$|a single jump of length 6 to the first instruction after $\mathit{CHECK}$.

In order to generate assignments we use the snippet 
\[
\mathit{NEXT}= \mathit{NEXT}_1; \cdots ;\mathit{NEXT}_k
\]
where
\[
\mathit{NEXT}_i= -\aux{i}.\mathtt{get};\#3;\aux{i}.\mathtt{set}\!\!:\!\!\fa;\#5; \aux{i}.\mathtt{set}\!\!:\!\!\tr
\]
for $1\leq i<k$ and 
\[
\mathit{NEXT}_k=-\aux{k}.\mathtt{get};\#3;\aux{k}.\mathtt{set}\!\!:\!\!\fa;\#3; \aux{k}.\mathtt{set}\!\!:\!\!\tr; \ferm
\]
Observe that, starting with the assignment $\oplus_{i=1}^{k} \aux{i}.B(\tr)$, repeated execution of $\mathit{NEXT}$ generates all possible assignments. After the last assignment $\oplus_{i=1}^{k} \aux{i}.B(\fa)$, all values are set back to $\tr$ and the generator terminates with reply $\fa$.

We combine the checks and the assignment generator and define
\[
I_k=\mathit{CHECK};\mathit{NEXT};\backslash \# (72k^3+5k).
\]
Then 
\[
(|I_k|/\oplus_{i=1}^{k} \aux{i}.B(\tr))! \oplus_{i=1}^{8k^3}\invar{i}.B(b_i) = 
\begin{cases}
\tr&\text{ if $\psi_{\langle b_i \rangle}$ is satisfiable},\\
\fa & \text{ otherwise.}
\end{cases}
\]
Thus $I_k$ computes $3SAT(k)$ using $k$ auxiliary registers.
Since $I_k$ has $72k^3 + 5k +1$ instructions, we may conclude that  $3SAT \in \mathcal{IS}_P(A)$. \hfill $\Box$
\begin{corollary}
If $NP\not \subseteq\mathrm{P/poly}$ then \(\mathcal{IS}^\mathit{lf}_P(A) \subsetneq \mathcal{IS}_P(A)\).
\end{corollary}
{\bf Proof:} Suppose $\mathit{3SAT}\in \mathcal{IS}^{lf}_P$, then $\mathit{3SAT}\in \mathrm{P/poly}$ by Theorem \ref{beneden} and hence 
$NP\subseteq\mathrm{P/poly}$.\hfill $\Box$

\section{Conclusion}
Program algebra is a setting suited for investigating instruction sequences. In this setting, we have shown that each partial Boolean function can be computed by an instruction sequence without the use of auxiliary registers or backward jumps. Hence backward jumps do not contribute to the expressiveness of instruction sequences. However, instruction sequences can be significantly shorter when backward jumps and auxiliary registers are permitted. Thus, semantically we can do without backward jumps. However, if we want to avoid an explosion of the length of instruction sequences, then backward jumps are essential. It remains an open problem whether the third inclusion in Corollary~\ref{hierarchy} is proper.

\end{document}